\documentclass[sigconf]{acmart}
\usepackage{enumitem}
\usepackage{multirow}
\usepackage{multicol}
\usepackage{booktabs}
\usepackage{url}
\usepackage{mathrsfs}
\usepackage{caption}
\usepackage{subfigure}

\copyrightyear{2022}
\acmYear{2022}
\setcopyright{acmcopyright}
\acmConference[SIGIR '22] {Proceedings of the 45th International ACM SIGIR Conference on Research and Development in Information Retrieval}{July 11--15, 2022}{Madrid, Spain.}
\acmBooktitle{Proceedings of the 45th International ACM SIGIR Conference on Research and Development in Information Retrieval (SIGIR '22), July 11--15, 2022, Madrid, Spain}
\acmPrice{15.00}
\acmISBN{978-1-4503-8732-3/22/07}
\acmDOI{10.1145/3477495.3531922}

\begin{document}
\fancyhead{}

\title{ReLoop: A Self-Correction Continual Learning Loop \\for Recommender Systems}

\author{Guohao Cai$^{1\dagger}$,$\:$ 
Jieming Zhu$^{1\dagger}$,$\:$ 
Quanyu Dai$^1$,$\:$ 
Zhenhua Dong$^{1\ast}$,$\:$ 
Xiuqiang He$^1$ \newline
Ruiming Tang$^1$,$\:$ 
Rui Zhang$^{2\ast}$ \vspace{1ex}}
\affiliation{%
  \institution{\textsuperscript{1}Huawei Noah's Ark Lab, Shenzhen \country{China} $\:$ $\:$  \textsuperscript{2}www.ruizhang.info}
  \texttt{\{caiguohao1,jamie.zhu,daiquanyu,dongzhenhua,tangruiming\}@huawei.com, rayteam@yeah.net}\vspace{2ex}
}

\thanks{$^\dagger$ Both authors contributed equally to this work}
\thanks{$^\ast$ Corresponding authors}

\renewcommand{\authors}{Guohao Cai, Jieming Zhu, Quanyu Dai, Zhenhua Dong, Xiuqiang He, Ruiming Tang, Rui Zhang}

\begin{abstract}
Deep learning-based recommendation has become a widely adopted technique in various online applications. Typically, a deployed model undergoes frequent re-training to capture users' dynamic behaviors from newly collected interaction logs. However, the current model training process only acquires users' feedbacks as labels, but fail to take into account the errors made in previous recommendations. Inspired by the intuition that humans usually reflect and learn from mistakes, in this paper, we attempt to build a self-correction learning loop (dubbed ReLoop) for recommender systems. In particular, a new customized loss is employed to encourage every new model version to reduce prediction errors over the previous model version during training. Our ReLoop learning framework enables a continual self-correction process in the long run and thus is expected to obtain better performance over existing training strategies. Both offline experiments and an online A/B test have been conducted to validate the effectiveness of ReLoop.
\end{abstract}

\keywords{Recommender System, CTR prediction, Self-Correction}

\maketitle
\section{Introduction}

Nowadays, recommender systems play an indispensable role in our daily life. With the ever-growing volume of online information, recommender systems have been widely adopted to help users discover items of their interests in various applications, such as e-commerce, news feeds, music apps, movie websites, and so on. The quality of recommendation could not only influences user experience, but also has a direct impact on platform revenues. As a consequence, the research of recommender systems has received much attention from both academia and industry in recent years~\cite{DeepSurvey}.



Model deployment in an industrial recommender system usually comprises two interleaved processes: offline (incremental) training and online inference. The deployed model needs to undergo frequent re-training to capture time-varying user behaviors from newly collected interaction logs. This forms the conventional learning loop: \textit{model training}$\rightarrow$\textit{online inference}$\rightarrow$\textit{feedback data}$\rightarrow$\textit{model training} $\rightarrow$ $\cdots$. First, after offline training, the model is deployed online to serve user requests. Then, new feedback data are collected as input (e.g., clicked user-item pairs as positive samples and unclicked ones as negative samples) for the next round of model training, which in turn delivers a new model version for online inference. Such a training process only acquires users' feedbacks (e.g., clicks) as supervision signals, but dose not take into account the errors made in previous model versions. This wastes the opportunity of accumulating knowledge from the past recommendation experience for self-correction. 








As an analogy, humans are taught to learn from mistakes. Taking class learning for example, students need to review their past wrong question collections regularly to achieve better grades in the next exam. This inspires us to think about how to reflect and learn from the past errors for recommender systems, which is an interesting yet unexplored research problem. Existing studies focus mostly on designing better models to learn from the historical data, e.g., via feature interactions~\cite{DeepFM,DCN_V2}, user behavior modeling~\cite{DIN,DIEN,UBR} and multi-task learning~\cite{ESMM, MMoE}. The most related work in this direction includes knowledge distillation~\cite{Self_KD1, Self_KD2} and continual learning~\cite{CLSurvey}. Knowledge distillation could be applied to mimicing the behaviors of previous models, while continual learning techniques aim to address the catastrophic forgetting issue during continuous model training. Yet, neither of them focuses on developing the self-correction ability of models that could continuously reflect on their past errors and improve themselves.



\begin{figure*}[!thbp]
	\centering
	\includegraphics[scale=0.38]{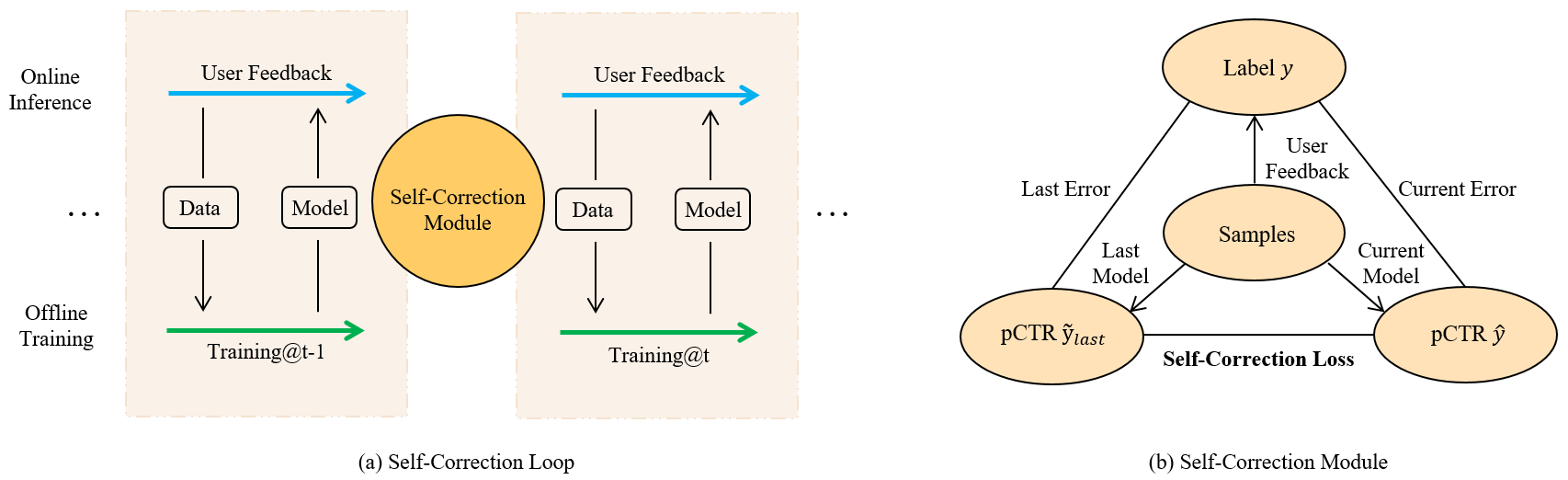}
	\caption{
	(a) illustrate two sequential training procedures, which are connected by the self-correction module; (b) details the self-correction module, where errors are defined from model's prediction score and users' feedback label. We expect the training process could reduce these errors for better online serving. 
	}
	\label{fig:framework}
\end{figure*}

In this work, we propose a novel learning framework named ReLoop, with the goal of building a self-correction learning loop for recommender systems. There is a large design space to explore for our ReLoop framework: for example, using sample-wise testing errors, list-wise ranking errors, or even online business metrics (e.g., clicks, revenues) of a recommendation list as the error measure. As an initial attempt in this work, we opt for a simple yet effective design choice and demonstrate its use in the click-through rate (CTR) prediction task. In the training loop of recommender systems, data samples are continuously collected from the deployed model online, and thus we could obtain both the click labels from users and the corresponding prediction errors from the previous models. Our key insight is that a new model should be forced to make smaller prediction errors than the previous model versions on the same data samples, and thus achieves the ability of self correction. With this insight, we choose sample-wise prediction errors (i.e., gaps between prediction scores and ground truths) as the error measure and further propose a customized loss function to force every new model version to reduce prediction errors over the previous model version during training. Our ReLoop framework enables a continual self-correction process in the long run and thus is expected to obtain better performance than existing training strategies. The framework is generic and model-agnostic because it is applicable to the training of any existing recommendation model.





To validate the effectiveness of ReLoop, we conduct extensive experiments on both public benchmark datasets and a private industrial dataset. The results from both offline experimental results and an online A/B test show consistent improvements over the baselines when applying our ReLoop learning framework. The main contributions of this paper are summarized as follows:
\begin{itemize}
    \item We identify the importance of reflecting on past errors, and propose a generic ReLoop learning framework for continuous self-correction in recommender systems.
    \item We design a simple and effective self-correction module based on a customized loss, which requires nearly no engineering work in the online serving system.
    \item Both offline experiments and an online A/B test have been conducted to demonstrate the simplicity and effectiveness of our ReLoop framework.
\end{itemize}

\section{ReLoop Framework}

In this section, we first describe the training loop in industrial recommender systems. Then we define the errors made in the loop, based on which a self-correction module is introduced. Finally, we integrate the self-correction module into the CTR prediction learning paradigm and compare it with the knowledge distillation method. 


\subsection{Training Loop In Production}
The training loop of a recommendation model is illustrated in Figure \ref{fig:framework}(a). Firstly, training data is collected from user implicit feedback on the exposed items, i.e., clicked items as positive samples and unclicked items as negative samples. Secondly, a ranking model is trained using a sliding window of such data (w.r.t. both offline/incremental training modes). Then, the new model will be updated to the online serving worker. During the online inference stage, candidates are evaluated by the model and displayed to users according to the ranking of prediction scores. Finally, online exposure and click events will be recorded in the user behavior log, which in turn triggers a new period of training. This forms the training loop for recommender systems. Nevertheless, we can see that the connection between adjacent training procedures are very loose, since each one trains its own model independently. In other words, current training procedure does not fully utilize information from previous procedure, and ignores the errors made in  previous predictions. To alleviate this problem, we introduce a novel self-correction module to learn from previous mistakes.

\subsection{Self-Correction Module}
Figure \ref{fig:framework}(b) presents the design details of the proposed self-correction module. Each sample has a label $y$, which is generated from users' implicit feedback (e.g., clicks). The predicted click-through rate (pCTR) could be recorded during model inference, which represents the probability of a user clicking on an item. Intuitively, we define the gap between a sample's pCTR and its ground truth label as the error made, so we could get the last error from the last pCTR $\widetilde{y}_{last}$ for each sample. As for the current error, it denotes the gap between the current prediction $\hat{y}$ and the ground truth label during training. Inspired by the fact that human beings learn from mistakes, we  expect the  model at time $t$ performs better than the last $t-1$ version. For the 
purpose of building such an evolving system, we design a novel self-correction loss, which forces the training process get lower error than the previous one. For clarity, we take a positive sample ($y=1$) for example:
\begin{equation}
  y - \hat{y}   \leq  y - \widetilde{y}_{last}\quad \Leftrightarrow \quad\widetilde{y}_{last} - \hat{y}  \leq 0
\end{equation}
where $y$ is the label, $\hat{y}$ is the predicted CTR in current training step, and $\widetilde{y}_{last}$ is the predicted CTR from last version of the model.
In order to reduce the error from the past and guide the training process to a better direction, we propose to  penalize those samples on which it performs worse than the last version using the following loss:
\begin{equation}
\mathcal{L}_{sc} = y \max(\widetilde{y}_{last} - \hat{y}, 0) + (1-y) \max(\hat{y} - \widetilde{y}_{last}, 0)~.
\label{eq:reloop} 
\end{equation}
The loss $\mathcal{L}_{sc}$ is an extension from the widely used hinge loss. Note that the last pCTR $\widetilde{y}_{last}$ could be obtained from previous model, or directly collected from the online inference server. In this way, samples incorrectly predicted by the previous model could be emphasized when the current model does not behavior better than the previous version.


\subsection{Training Strategy}
The common objective function for CTR prediction task is the binary cross-entropy loss as follows:
\begin{equation}
\mathcal{L}_{ce} = -y \log\hat{y} - (1-y) \log(1-\hat{y} )
\label{eq:ce}
\end{equation}
Then our ReLoop loss function is formulated as a sum of both terms:
\begin{equation}
 \mathcal{L}= \alpha\mathcal{L}_{sc} + (1-\alpha)\mathcal{L}_{ce}
\label{eq:aug}
\end{equation}
where $\alpha\in[0,1]$ is the hyper-parameter to adjust the importance of self-correction loss and cross-entropy loss. Thanks to the design for  reflecting on past errors, our proposed ReLoop framework enables the CTR prediction task to be optimized in a self-correction manner. 

\subsection{Comparison with Knowledge Distillation}\label{subsec:kd}
\begin{figure}[htbp]
    \centering
    \subfigure[$y=1,\widetilde{y}_{last}=0.8$]{
        \includegraphics[width=4cm]{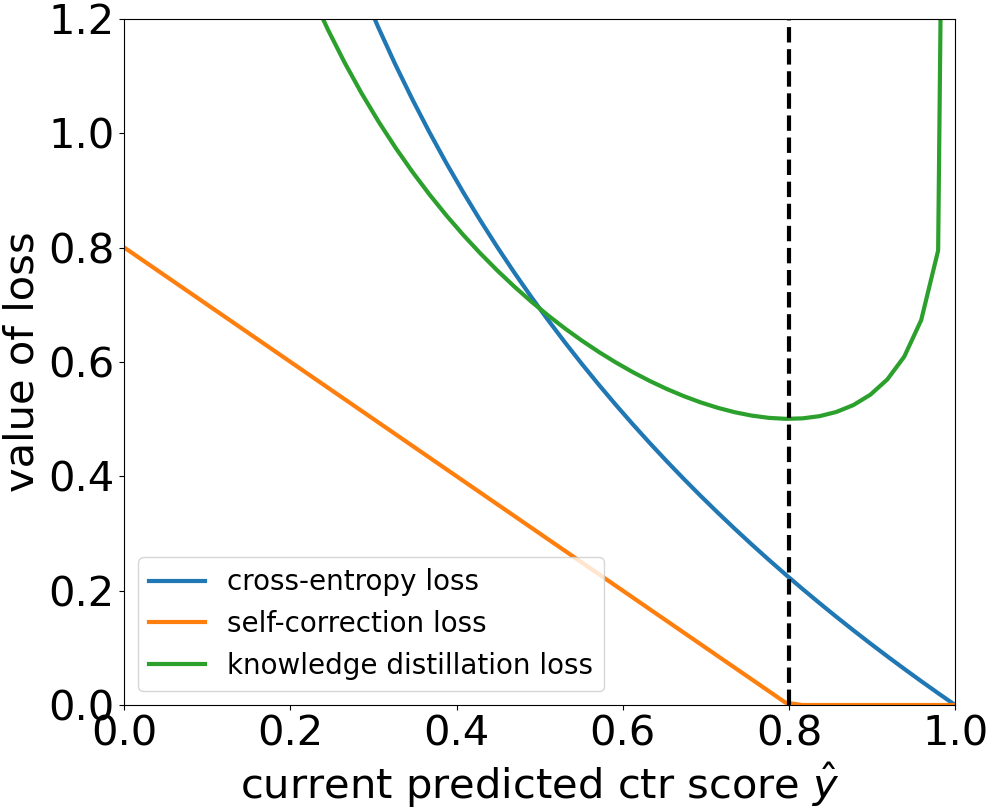}
    }
    \subfigure[$y=0,\widetilde{y}_{last}=0.3$]{
	\includegraphics[width=4cm]{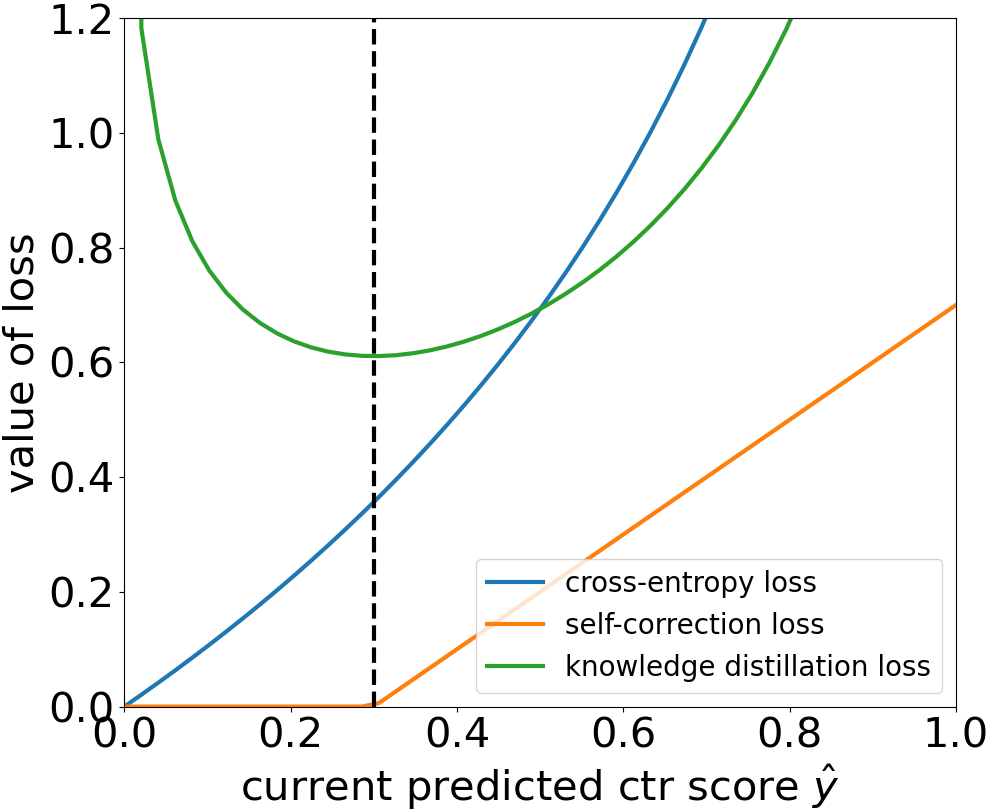}
    }
    \caption{The Loss functions of cross-entropy loss, knowledge distillation loss and our proposed self-correction loss.}
    \label{fig:loss}
\end{figure}
{Our design bears a similarity to the technique of knowledge distillation~\cite{KD}, which could be formulated as follows:
\begin{equation}
\mathcal{L}_{kd} = -\widetilde{y}_{last} \log\hat{y} - (1-\widetilde{y}_{last}) \log(1-\hat{y} )
\label{eq:kd}
\end{equation}
The main difference is that knowledge distillation is designed to mimic the behaviors of a teacher model, while our self-correction learning module is proposed to learn from past errors of the previous model. It is worth noting that teachers could make mistakes too. Take Figure \ref{fig:loss} (a) for example, given a positive sample, the predicted score from the previous model is $\widetilde{y}_{last}=0.8$. If current prediction $\hat{y}$ is lower than $\widetilde{y}_{last}$, namely performing worse than the previous version, the basic $\mathcal{L}_{ce}$ loss will be assisted by the self-correction loss $\mathcal{L}_{sc}$. On the other hand, if $\hat{y}$ has already been higher than $\widetilde{y}_{last}$, namely performing better than before, we do not impose extra penalty on the model.
}

\section{Experiments}

In this section, we first conduct extensive offline experiments on four public datasets. Then, we deploy our ReLoop framework in an application of one industrial Newsfeed product, where hundreds of millions of users browse news and videos everyday. 
We aim to answer the following research questions:
\\
\textbf{RQ1}: How does our proposed method ReLoop perform against the state-of-the-art methods?
\\
\textbf{RQ2}: How does the hyper-parameter influence the performance?
\\
\textbf{RQ3}: How dose ReLoop perform in online A/B tests?

\subsection{Evaluation on Public Datasets}
\subsubsection{Experiment Setup}
We conduct experiments on four public datasets following previous works \cite{DBLP:conf/aaai/ChengSH20,DBLP:conf/sigir/0001C17,DBLP:journals/tois/QuFZTNGYH19}, and on a large-scale industrial dataset.
\begin{itemize}[leftmargin=*]
\item \textbf{Criteo} is a famous benchmarking dataset for CTR prediction tasks. It contains 45M instances, and each instance consists of 13 numerical feature fields and 26 categorical fields.
\item \textbf{Avazu} is a mobile advertisements dataset. It contains 40M instances with 22 feature fields including advertisement attributes and user features.
\item \textbf{MovieLens} dataset consists of users' tagging records on movies. It contains 2M instances and focus on the personalized tag recommendation.
\item \textbf{Frappe} dataset is collected from a context-aware app discovery tool. It contains 288k instances and 10 fields for each input.
\item \textbf{Production} dataset is consisted of more than 500M instances and each one has over 100 feature fields, sampled from an industrial news feed product.
\end{itemize}
For the public datasets, we split the instances by 8:1:1 for training, validation and testing, following the experiments setting by \cite{DBLP:conf/aaai/ChengSH20}. Since these public datasets do not contain historical user behavior sequences, user behavior based models are not suitable for comparison. To get last prediction score for public datasets, we use 90\% of training samples, mimicking the online setting, to train the previous model version and then infer on the whole training set to generate $\widetilde{y}_{last}$ values . 


\begin{table}[!thp]
	\centering
	\setlength{\tabcolsep}{0.5pt}
	\small
	\caption{Comparison of competing methods and ReLoop on Four public datasets.}
	\begin{tabular}{ccccccccc}
	\toprule
\multirow{2}{*}{Models} & \multicolumn{2}{c}{Criteo} & \multicolumn{2}{c}{Avazu} &\multicolumn{2}{c}{MovieLens} &\multicolumn{2}{c}{Frappe} \\
&AUC &Logloss &AUC &Logloss &AUC &Logloss &AUC &Logloss\\
\midrule
LR  & 0.7858    & 0.4636    & 0.7313    & 0.4065    & 0.9215    & 0.3080  & 0.9329    & 0.2860     \\

FM  &0.7933    &0.4574    &0.7496    &0.3740    &0.9388    &0.2797    &0.9641    &0.2143        \\


NFM   &0.7968    &0.4537    &0.7531    &0.3761    &0.9441    &0.3004    &0.9727    &0.2079     \\


IPNN      &0.8026    &0.4509    &0.7526    &0.3737    &0.9469    &0.2792    &0.9735    &0.2012     \\


Wide$\&$Deep     &0.8062    &0.4453    &0.7529    &0.3744    &0.9381    &0.3310    &0.9728    &0.2038    \\
DCN       &0.8059    &0.4463    &0.7550    &0.3721    &0.9419    &0.2791    &0.9402    &0.2808    \\
DeepFM      &0.8025    &0.4501    &0.7535    &0.3742    &0.9424    &0.3131    &0.9719    &0.2108      \\

xDeepFM      &0.8070    &0.4443    &0.7535    &0.3737    &0.9448    &0.2717    &0.9738    &0.2098     \\

FmFM      & 0.8056     & 0.4462   &  0.7603  & \underline{0.3685}  & 0.9465  & 0.2714   & 0.9749   & 0.2004  \\

AFN+     &0.8074    &0.4451    &0.7555    &0.3718    &0.9500    &0.2585    &0.9783    &0.1762 \\

\midrule
KD+DCN    &\underline{0.8136}   &\underline{0.4473}   &\underline{0.7653}   &0.3906  &0.9648  &\underline{0.2164}   &\underline{0.9837}  &0.1672                         \\
KD+DeepFM  &0.8135   &0.4462   &0.7652   &0.3690  &\underline{0.9650}  &0.2183   &0.9803  &\underline{0.1552}                         \\
\midrule
RLP+DCN  &0.8138 & \textbf{0.4381}  & 0.7648 & \textbf{0.3667} & \textbf{0.9693}  & 0.2189 &0.9843 &0.1660     \\ 

RLP+DeepFM  &\textbf{0.8139}   &0.4382   &\textbf{0.7659}   &0.3750  &0.9687 &\textbf{0.2160} &\textbf{0.9848}  &\textbf{0.1549}                    \\

    \bottomrule
	\end{tabular}
	\label{tab:offline}
\end{table}

\textbf{Baselines.} We compare our ReLoop model with many state-of-the-art approaches including LR~\cite{FTRL}, FM~\cite{FM}, NFM~\cite{NFM}, PNN~\cite{PNN}, Deep$\&$Cross~\cite{DeepCross}, Wide$\&$Deep~\cite{WideDeep}, DeepFM~\cite{DeepFM}, xDeepFM~\cite{xDeepFM}, FmFM~\cite{DBLP:conf/www/SunPZF21}, AFN+~\cite{DBLP:conf/aaai/ChengSH20}, Distill~\cite{KD}.

\subsubsection{Experiment results}
The experiments results are shown in Table \ref{tab:offline}. We take Area under ROC curve (AUC) and Logloss as offline metrics. We utilize two representative methods, namely DCN and DeepFM as backbone for our framework. From this table, we can see that methods based on ReLoop outperform all the other baselines by a large margin, demonstrating the superiority and generalization to different datasets. ReLoop can be regarded as plug-in component for improving CTR models performance. Compared to original backbone models, our framework can polish up consistently and significantly. Specifically, It achieves AUC 1.42\%, 1.65\%, 2.91\% and 4.69\% improvement relative to the backbone models, which prove the effectiveness of our novel self-correction module. In addition, knowledge distillation technique can also promote the accuracy and become the best baseline on almost all datasets.

\subsection{Evaluation in Production}
\begin{figure}[!thb]
	\centering\small
	\includegraphics[scale=0.28]{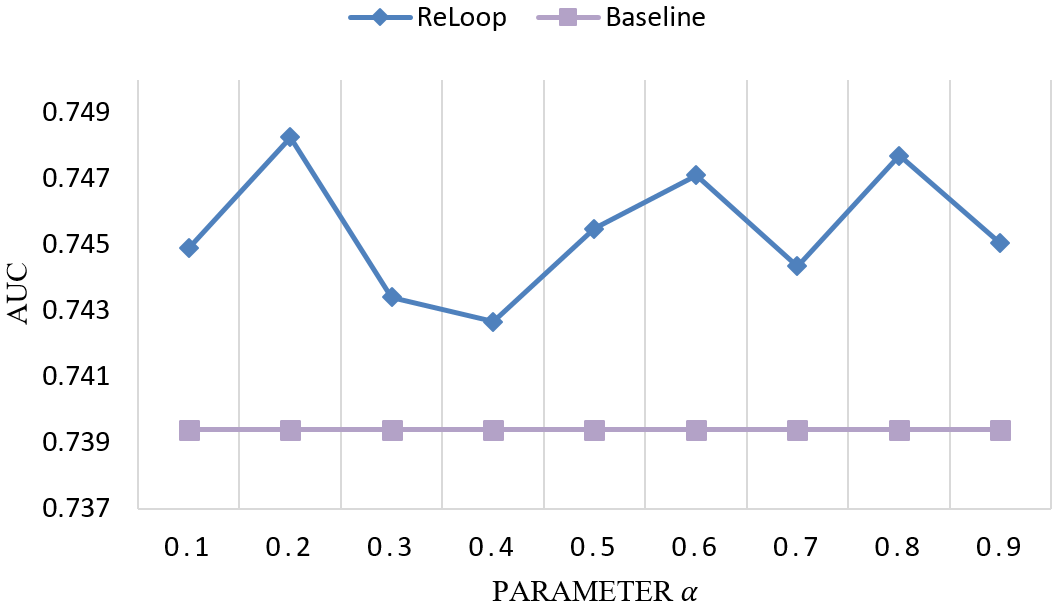}
	\caption{The comparison between baseline and ReLoop on different value of parameter $\alpha$ in terms of AUC metric.}
	\label{fig:auc}
\end{figure}

Inspired by the promising improvement of ReLoop framework on public datasets, we deploy our method to a real world NewsFeeds recommender system. Firstly, the offline experiments are conducted, and results are shown in Table \ref{tab:distill}. We can see that ReLoop outperform baseline and knowledge distillation method on both AUC and Logloss metrics. It is worth noting that our framework introduces little cost on training and no cost on inference, making it convenient to be deployed on existing system. Besides, we conduct sensitive experiments in terms of  hyper-parameter $\alpha$ from equation \ref{eq:aug}, shown in the Figure \ref{fig:auc}. The purple line is the baseline method which does not adopt the self-correction loss and we can see that ReLoop framework perform better consistently.
\begin{table}[!thb]
	\centering
	\setlength{\tabcolsep}{2pt}
	\caption{Comparison between knowledge distillation and ReLoop on production dataset.} 
	\begin{tabular}{c|cc|cc|c}
	\toprule
    & AUC  & Improv.  & Logloss & Improv. & Time cost\\
    \hline
	Baseline & 0.7394 & -   & 0.2959  & - & - \\ 
	KD+Baseline  & 0.7458 & +0.87\%    &0.2743  & +7.30\% & +1.8\% \\
	RLP+Baseline &0.7483  & +1.20\%  &0.2313 & +21.83\% & +1.8\%   \\
	\bottomrule
	\end{tabular}
	\label{tab:distill}
\end{table}

\begin{table}[!thb]
	\centering
	\caption{The improvement of our ReLoop method compared to baseline in one-week online A/B tests.}
	\begin{tabular}{cccc}
	\toprule
	Day 1 & Day 2 & Day 3 & Day 4 \\
	\midrule
	+1.38\% & +2.01\% & +1.34\% & +0.64\%   \\ 
	\midrule
	Day 5 & Day 6 & Day 7 & Average \\
	\midrule
	+2.61\% & +1.13\% & +1.09\% & +1.46\%   \\ 
	\bottomrule
	\end{tabular}
	\label{tab:online}
\end{table}

For online A/B tests, we split the users into two groups, each of which has more than 300 thousands activate users. The first group receives recommendation results generated by a prediction model without ReLoop approach, which has been used by the current product; the second one receives the recommended news by the same backbone with ReLoop approach. We firstly conduct one week A/A tests to record the minor difference between two user groups and take CTR as the evaluation metric. During one week online A/B tests, our ReLoop method is consistently better than the baseline model, where CTR improvement achieves about 1.46\% on average (A/B - AA), detailed improvements are shown in Table \ref{tab:online}, which verify the effectiveness of our method. Deploying ReLoop to an existing CTR model is convenient and requires nearly no engineering work in the online serving system.




\section{Related Work}
In this section, we review the related work from three aspects: \textit{CTR prediction}, \textit{continual learning}, and \textit{knowledge distillation}.

\textbf{CTR Prediction}. The CTR (click-through rate) prediction task enables the ranking of a small set of candidate items by modeling fine-grained features of users, items and contexts. This task usually involves large-scale highly sparse data, which consist of multi-fields of numerical or categorical features.  
Due to the huge research and economic value of this task, many research efforts have been made in this direction. Typical models of CTR prediction range from simple logistic regression (LR) models~\cite{FTRL}, factorization machines (FM)~\cite{FM}, to deep neural networks (DNN)~\cite{WideDeep,DeepFM}. Readers could refer to the recent benchmarking work on CTR prediction and recommendation~\cite{FuxiCTR, BARS} for a more comprehensive view. 


In summary, current research on CTR prediction focuses mostly on the design of effective model structures, including feature interactions~\cite{DeepFM, DCN_V2}, sequential user behaviour modeling~\cite{DIN, DIEN}, multi-task learning~\cite{ESMM, MMoE}, multi-scenario prediction~\cite{STAR}, automated architecture search~\cite{auto-fi-autofis}, and so on. Nevertheless, the research community focuses less on the learning process in a continuous training loop, which is equally important in practice. Our work optimizes the learning loop through self-correction from past errors.

\textbf{Continual Learning}. Continual learning~\cite{CLSurvey}, as a lifelong learning concept, has received much attention in recent years. In a typical setting of continual learning, a model is set to learn different tasks sequentially without forgetting knowledge obtained from the preceding tasks. Therefore, most research focuses on mitigating the catastrophic forgetting issue and maximally retaining its accuracy on the old tasks that have been trained. Instead, in the continual learning process of recommender systems (e.g.,~\cite{Kraken}), the model is periodically trained on the same task with new data that may have distribution shift to a certain degree. Meanwhile, we only care about the model performance on the new data. Yet, the similarity lies in that our work also intends to retain the past knowledge in the continuous training process for model improvement. This new continual learning setting in recommender systems may deserve more attention and exploration. A concurrent work~\cite{CL_CTR} also applies continual learning for CTR prediction. In contrast, we focus on the self-correction ability in continual learning. 

\textbf{Knowledge Distillation}. The idea of knowledge distillation (KD)~\cite{KD}, which aims to distill the dark knowledge from a teacher model to a student model, has been widely applied in model compression and knowledge transfer~\cite{KD_survey}. Some recent work also applies KD to recommendation tasks, e.g., CTR prediction~\cite{EnsembleCTR}, topology distillation~\cite{TopologyKD}, and multi-task transfer~\cite{MultiTaskKD}. Instead, our work is mostly related to self-distillation~\cite{Self_KD1, Self_KD2}, which proposes to improve the generability of a model through previous model checkpoints or predictions. They regard the previous model as a teacher and take the prediction scores from teacher as soft labels to guide the current training process. However, \textit{teachers are not always correct}, and these methods fail to learn from the mistakes made in previous models. We make a distinction between our ReLoop loss and the KD loss in Figure \ref{fig:loss}.

\section{CONCLUSION and Future Work}
In this paper, we propose a generic learning framework, named ReLoop, for recommender systems, which can build a self-correction learning loop. Inspired by the learning paradigm of human beings, we introduce a novel self-correction module that takes into account the errors made in previous recommendations. In order to actively reflect on past errors, we design a customized loss that penalizes those samples on which the model performs worse than the past. Meanwhile, the proposed ReLoop framework applies to any existing model in recommender systems. Extensive experiments on four popular public datasets and one large-scale real industrial dataset have been conducted, and the performance results show the superiority of our framework than other state-of-the-art methods. In addition, we have deployed the framework on an industrial Newsfeed scenario and the online A/B test demonstrate the effectiveness of ReLoop method. In addition, there are some potential directions for future exploration. Examples include more fine-grained definition of errors, such as list-wise utility or session-based revenue, and more effective implementation (beyond loss) of the self-correction module. We hope our
work could inspire more research in this direction.

\section{acknowledgement}
We thank MindSpore\footnote{https://www.mindspore.cn/} for the partial support of this work, which is a new deep learning computing framework.

\bibliographystyle{ACM-Reference-Format}
\bibliography{reloop}
\appendix
\end{document}